 \documentclass[final,3p,times, 11pt]{elsarticle}
\usepackage{amssymb}
 \biboptions{sort&compress}

\begin{document}

\begin{frontmatter}

\begin{flushright}
LAL 11-109
\end{flushright}

%% Title, authors and addresses

%% use the tnoteref command within \title for footnotes;
%% use the tnotetext command for the associated footnote;
%% use the fnref command within \author or \address for footnotes;
%% use the fntext command for the associated footnote;
%% use the corref command within \author for corresponding author footnotes;
%% use the cortext command for the associated footnote;
%% use the ead command for the email address,
%% and the form \ead[url] for the home page:
%%
%% \title{Title\tnoteref{label1}}
%% \tnotetext[label1]{}
%% \author{Name\corref{cor1}\fnref{label2}}
%% \ead{email address}
%% \ead[url]{home page}
%% \fntext[label2]{}
%% \cortext[cor1]{}
%% \address{Address\fnref{label3}}
%% \fntext[label3]{}

\title{Impact on the Higgs Production Cross Section and Decay 
Branching Fractions of Heavy Quarks and Leptons in a Fourth Generation Model}

%% use optional labels to link authors explicitly to addresses:
%% \author[label1,label2]{<author name>}
%% \address[label1]{<address>}
%% \address[label2]{<address>}

\author[a,b]{X. Ruan}
\author[a]{Z. Zhang\corref{cor}}
\cortext[cor]{Corresponding author}
\ead{zhangzq@lal.in2p3.fr}

\address[a]{Laboratoire de l'Acc{\'e}l{\'e}rateur Lin{\'e}aire,
             Universit\'e Paris-Sud 11, IN2P3/CNRS, Orsay, France}
\address[b]{Institute of High Energy Physics, Chinese Academy of Sciences, 
Beijing, China}

\begin{abstract}
%% Text of abstract
In a fourth generation model with heavy quarks, the production cross
  section of the Higgs boson in the gluon-gluon fusion process is significantly 
  increased due to additional quark loops. In a similar way, 
  the partial decay width of the decay channels
  $H\rightarrow gg, \gamma \gamma$ and $\gamma Z$ is modified.
  These changes and their impact on the Higgs search are discussed. 
\end{abstract}

\begin{keyword}
%% keywords here, in the form: keyword \sep keyword
Fourth generation \sep Higgs production \sep Higgs decays
%% MSC codes here, in the form: \MSC code \sep code
%% or \MSC[2008] code \sep code (2000 is the default)

\end{keyword}

\end{frontmatter}

%%
%% Start line numbering here if you want
%%
% \linenumbers

%% main text
\section{Introduction}
\label{sec:intro}
The Standard Model (SM) are known to have three families of 
charged and neutral fermions. There is however no upper limit on the number 
of fermion families. 
A fourth family (SM4)~\cite{fhs99} could be in fact the key to many unsolved 
puzzles,
such as the hierarchies of the fermion mass spectrum including neutrino masses 
and mixing, electroweak symmetry breaking, baryogenesis,
and a variety of interesting phenomena in CP and flavor 
physics~\cite{hps01,holdom09,elr10,mc2010,chilr11}.

%A fourth generation is neither ruled out by the current
%precision electroweak data nor excluded by the direct searches.

Fourth family leptons and quarks have been searched for previously by the LEP 
and Tevatron experiments and now by the LHC experiments.
The most stringent lower mass limits at 95\% CL are~\cite{pdg10,limit_b4,limit_t4}
\begin{eqnarray}
&& m_{\nu_4}>80.5 - 101.5\,{\rm GeV}\,,\label{eq:nu4_limit}\\
&& m_{l_4}>100.8\,{\rm GeV}\,,\\
&& m_{b_4}>372\,{\rm GeV}\,,\\
&& m_{t_4}>335\,{\rm GeV}\,.
\end{eqnarray}
The mass bound on the heavy neutrino depends on the type of neutrino (Dirac or 
Majorana) and whether one considers a coupling of the heavy neutrino to 
$e^-, \mu^-$ or $\tau^-$. It should also be noted that assumptions about 
the coupling of the fourth family and the decay mode were made in deriving 
the quark mass limits. The limits can be weaker without these assumptions. 
On the other hand, the triviality bound from unitarity of 
the $t_4 t_4$ $S$-wave scattering~\cite{cfh79} indicates a 
maximum $t_4$ mass of around 500\,GeV, although this estimate is based on 
tree-level expressions.

With the presence of fourth family quarks, the dominant Higgs production 
process, the gluon-gluon fusion process, is further enhanced 
(see Sec.~\ref{sec:sigma} and also~\cite{kpst07,scis10}).
Using the large enhancement of $gg\rightarrow H\rightarrow WW$ in SM4, 
CDF, D0 and CMS have been able to exclude the SM4 Higgs boson at 95\% CL
for $131\le m_H\le 204$\,GeV~\cite{tevatron} and $144\le m_H\le 207$\,GeV~\cite{cms}.

In~\cite{rv10} (see also \cite{bfkks02,ks11}), it was pointed out that the mass
limit on $m_{\nu_4}$ (Eq.(\ref{eq:nu4_limit})) was derived 
when the mixing angle between the fourth family neutrino and at least
one of the neutrinos in the SM is assumed to be larger than $3\times 10^{-6}$.
For smaller mixing angles, $\nu_4$ is quasistable and the mass of $\nu_4$ is
bounded only from the analysis of $Z$ boson decay at 
$m_{\nu_4}>46.7$\,GeV~\cite{bnv03}. In this case, the decay mode 
$H\rightarrow \nu_4\nu_4$ becomes the dominant one at low Higgs mass and 
the Tevatron lower mass limits of 131\,GeV would be 
increased to 155\,GeV~\cite{rv10}. In the following, we will not consider 
this possibility.

\section{Enhancement of gluon-gluon fusion Higgs production cross 
section}\label{sec:sigma}

In the SM, the dominant Higgs production process is the gluon-gluon fusion 
process where gluons from colliding beams couple to a heavy quark loop 
from which the Higgs boson is emitted.
The cross section at the leading-order (LO) can be written as~\cite{bp97}
\begin{equation}
\sigma(pp\rightarrow HX)=\Gamma(H\rightarrow gg)\frac{\pi^2}{8m^3_H}\tau_H\int^1_{\tau_H}\frac{dx}{x}g(x,m^2_H)g(\tau_H/x,m^2_H)
\end{equation}
with
\begin{equation}
\Gamma(H\rightarrow gg)=\frac{G_Fm^3_H}{36\sqrt{2}\pi}\left[\frac{\alpha_s(m^2_H)}{\pi}\right]^2|I|^2\,,\hspace{5mm} I=\sum_q I_q
\end{equation}
and $g(x,Q^2)$ is the gluon distribution evaluated at $x$ and $Q^2$, $G_F$ ($\alpha_s$) is the Fermi (strong) coupling constant. The quantity $I_q$ is given 
in terms of $\lambda_q=m^2_q/m^2_H$:
\begin{equation}
I_q=3\left[2\lambda_q+\lambda_q(4\lambda_q-1)f(\lambda_q)\right]\,,
\end{equation}
where
\begin{equation}
{\displaystyle
f(\lambda_q)=\left\{\begin{array}{lr}\displaystyle-2\left(\sin^{-1}\frac{1}{2\sqrt{\lambda_q}}\right)^2\,,& \mbox{for $\displaystyle\lambda_q>\frac{1}{4}\,,$}\\\displaystyle
\frac{1}{2}\left(\ln\frac{\eta^+}{\eta^-}\right)^2-\frac{\pi^2}{2}-i\pi\ln\frac{\eta^+}{\eta^-}\,,& \mbox{for $\displaystyle\lambda_q<\frac{1}{4}$}\,,
\end{array}\right.}
\end{equation}
with $\eta^\pm=\frac{1}{2}\pm \sqrt{\frac{1}{4}-\lambda_q}$.
In the heavy quark $m_q$ limit, $\lambda_q\gg 1$, $I_q\rightarrow 1$, whereas in the
light quark $m_q$ limit $\lambda_q<\!\!< 1$, $I_q\rightarrow 0$. 
This is why in the SM, the top quark is by far the dominant contribution.

\begin{figure}
  \begin{center}
  \includegraphics[width=0.55\columnwidth]{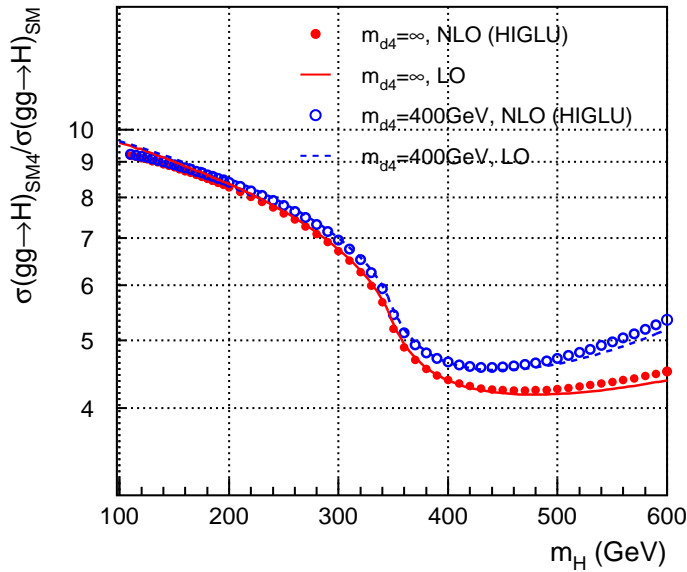}
  \end{center}
  \caption{The enhancement factor of the Higgs production cross section in a
fourth generation over that of the SM as a function of $m_H$ for two scenarios
with $m_{d_4}=10\,{\rm TeV}$ and $400\,{\rm GeV}$.}
  \label{fig:rsigma}
\end{figure}

In a fourth generation with two additional heavier quark $t_4$ and $d_4$, the
Higgs production cross section is enhanced with that of the SM by a factor
\begin{equation}
R^{\rm SM4/SM}_{\sigma(gg\rightarrow H)}\equiv \frac{\sigma(gg\rightarrow H)_{\rm SM4}}{\sigma(gg\rightarrow H)_{\rm SM}}=\frac{|I_b+I_t+I_{t_4}+I_{d_4}|^2}{|I_b+I_t|^2}\,.\label{eq:rsigma}
\end{equation}
The dependence as a function of the Higgs boson mass $m_H$ is shown in 
Fig.~\ref{fig:rsigma} for two different $m_{d_4}$ values of infinite
mass~\footnote{In practice, a value of 10\,TeV is chosen.} and 400\,GeV. 
The $t_4$ mass is fixed as~\cite{kpst07}
\begin{equation}
m_{t_4}=m_{d_4}+50+10\times\ln\left(\frac{m_H}{115\,[{\rm GeV}]}\right)\,,
\end{equation}
to be consistent with the constraint of electroweak precision data.
The maximum enhancement factor of about 9 is reached at the small Higgs boson 
mass $m_H$ value where $I_b\rightarrow 0$ and $I_t, I_{t_4}, I_{d_4}\rightarrow 1$.
At low $m_H$, the enhancement is independent of the quark mass value of 
the fourth generation. At higher Higgs mass values, the heavier the quark mass,
the smaller the enhancement factor. The heavy $m_{q_4}$ scenario may not be 
physical as when 
it is beyond about $500$\,GeV the weak interaction among heavy
particles becomes strong and perturbation theory breaks down.
However, since the enhancement is the smallest, the resulting exclusion
limits would be more conservative. This is the scenario used by CMS in their
recent publication~\cite{cms}. The other scenario with $m_{d_4}=400$\,GeV 
corresponds to one of the scenarios used by the Tevatron experiments~\cite{tevatron}.

In Fig.~\ref{fig:rsigma}, the enhancement based on LO cross sections is also
compared with the corresponding factor in NLO calculated with a 
modified {\sc HIGLU} program~\cite{higlu}~\footnote{Following the suggestion
of the author, the SM electroweak corrections are not applied (by setting
ELW=0 in the steering file higlu.in) as they are not valid for a fourth 
generation model.}. The cross sections $\sigma(gg\rightarrow H)_{\rm SM, SM4}$ 
are calculated for $pp$ collisions at 7\,TeV center-of-mass energy~\footnote{Whereas the cross section values depend strongly on the center-of-mass energy, we have checked that the ratio $R^{\rm SM4/SM}_{\sigma(gg\rightarrow H)}$ has essentially no dependence on it.}.
As far as the ratio $R^{\rm SM4/SM}_{\sigma(gg\rightarrow H)}$ is concerned, 
the difference between NLO and LO is small. This means that one may use
this ratio (e.g.\ in NLO) and precisely predicted SM cross section values in 
higher orders to derive the corresponding higher-order cross section in 
a fourth generation model:
\begin{equation}
\sigma_{\rm SM4}(gg\rightarrow H)=\sigma_{\rm SM}(gg\rightarrow H)\times R^{\rm SM4/SM}_{\sigma(gg\rightarrow H)}\,.
\end{equation}
Indeed, use this relation and take the NNLO cross section value $\sigma_{\rm SM}(gg\rightarrow H)=19.81$ \,pb at $m_H=110$\,GeV as an example (Table~1 
in~\cite{handbook}) and the corresponding enhancement factor of 9.223 
(see the linked web page below), 
we obtain 182.79\,pb for the low mass scenario. 
This derived cross section value $\sigma_{\rm SM4}(gg\rightarrow H)$ is in 
excellent agreement with the independent 
prediction of 182.51\,pb given in Table~1 in~\cite{abfhl11}.

\section{Modified Higgs decay branching fractions}

\begin{figure}
  \begin{center}
  \includegraphics[width=0.55\columnwidth]{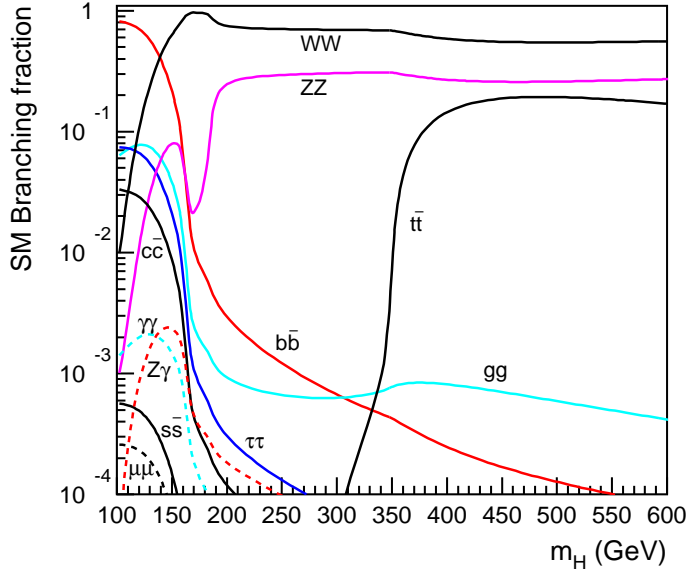}
  \end{center}
  \caption{Branching fractions of the SM Higgs decays calculated using
HDECAY.}
  \label{fig:br}
\end{figure}

The SM Higgs decay branching fractions calculated using HDECAY~\cite{hdecay} is 
displayed in Fig.~\ref{fig:br}.
The decay mode $H\rightarrow gg$ is the reverse of the gluon-gluon fusion
process $gg\rightarrow H$. The partial decay width of the decay mode 
$H\rightarrow gg$ in a fourth generation model with respect to that of the SM 
is thus enhanced by the same factor as the corresponding Higgs production 
cross section in the gluon-gluon fusion process (Eq.(\ref{eq:rsigma})).

The decay mode $H\rightarrow \gamma\gamma$ is similar to $H\rightarrow gg$ 
except that charged leptons, the $W$-boson and charged Higgs bosons also 
contribute to the loop. The partial decay width is~\cite{bp97}
\begin{equation}
\Gamma(H\rightarrow\gamma \gamma)=\frac{G_F m^3_H}{8\sqrt{2}\pi}\left(\frac{\alpha}{\pi}\right)^2|I|^2\,,
\end{equation}
where $\alpha$ is the fine-structure coupling constant.
The quark, lepton, $W$-boson and colorless charged scalar
contributions are~\cite{bp97}:
\begin{equation}
I=\sum_q Q^2_qI_q+\sum_l Q^2_lI_l+I_W+I_S
\end{equation}
where $Q_f$ denotes the charge of fermion $f$ in units of $e$ and
\begin{eqnarray}
&& I_q=3\left[2\lambda_q+\lambda_q(4\lambda_q-1)f(\lambda_q)\right]\,,\\
&& I_l=2\lambda_l+\lambda_l(4\lambda_l-1)f(\lambda_l)\,,\\
&& I_W=3\lambda_W(1-2\lambda_W)f(\lambda_W)-3\lambda_W-\frac{1}{2}\,,\\
&& I_S=-\lambda_S\left[1+2\lambda_S f(\lambda_S)\right]\,,
\end{eqnarray}
and $\lambda_i=m^2_i/m^2_H$.
In a fourth generation model, both $I_Q$ and $I_l$ terms receive additional
contributions from fourth generation quarks $q_4$ and lepton $l_4$. 

\begin{figure}
  \begin{center}
  \includegraphics[width=0.49\columnwidth]{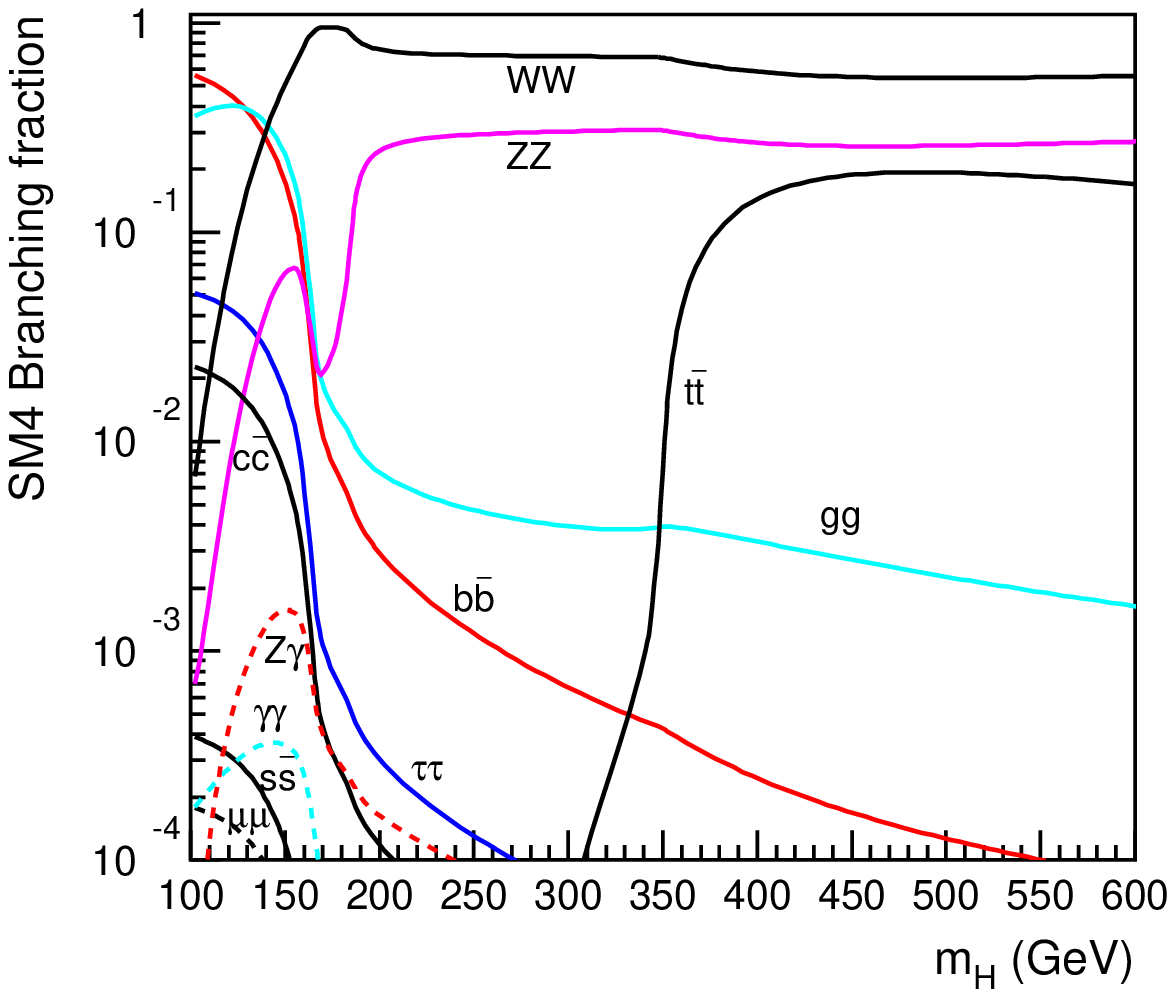}
  \includegraphics[width=0.49\columnwidth]{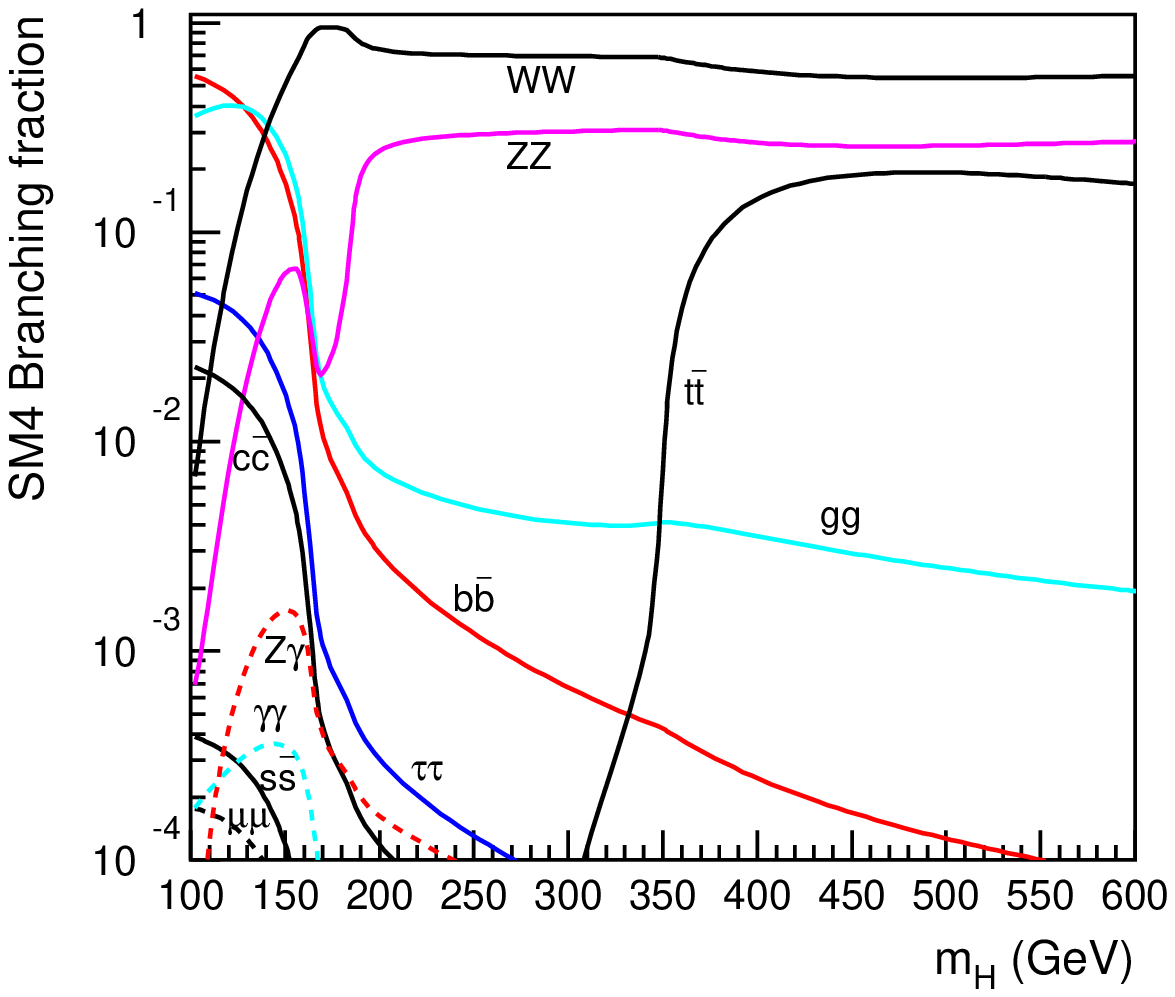}
  \end{center}
  \caption{Branching fractions of Higgs decays in a fourth generation model
with $m_{d_4}=m_{l_4}=10$\,TeV (left) and $m_{d_4}=m_{l_4}=400$\,GeV (right).}
  \label{fig:br_sm4}
\end{figure}

Finally the partial decay width $\Gamma(H\rightarrow \gamma Z)$ is also affected
by additional fourth generation quark loops. Therefore the branching fractions
(Fig.~\ref{fig:br_sm4}) in a fourth generation model look different from that
of the SM in particular for $gg$, $\gamma\gamma$ and $\gamma Z$ modes at
low Higgs mass values.

In Fig.~\ref{fig:br_sm4}, the two mass scenarios are compared. The $l_4$ mass
is further assumed to be the same as that of $d_4$. But as far as these
fourth generation quark and lepton masses are heavy enough ($>m_H/2$),
the difference in branching fractions is hardly visible.
The branching fractions in a fourth generation model are calculated without
applying the electroweak and higher-order QCD corrections as they do not apply 
to the fourth generation quarks.

\begin{figure}
  \begin{center}
  \includegraphics[width=0.49\columnwidth]{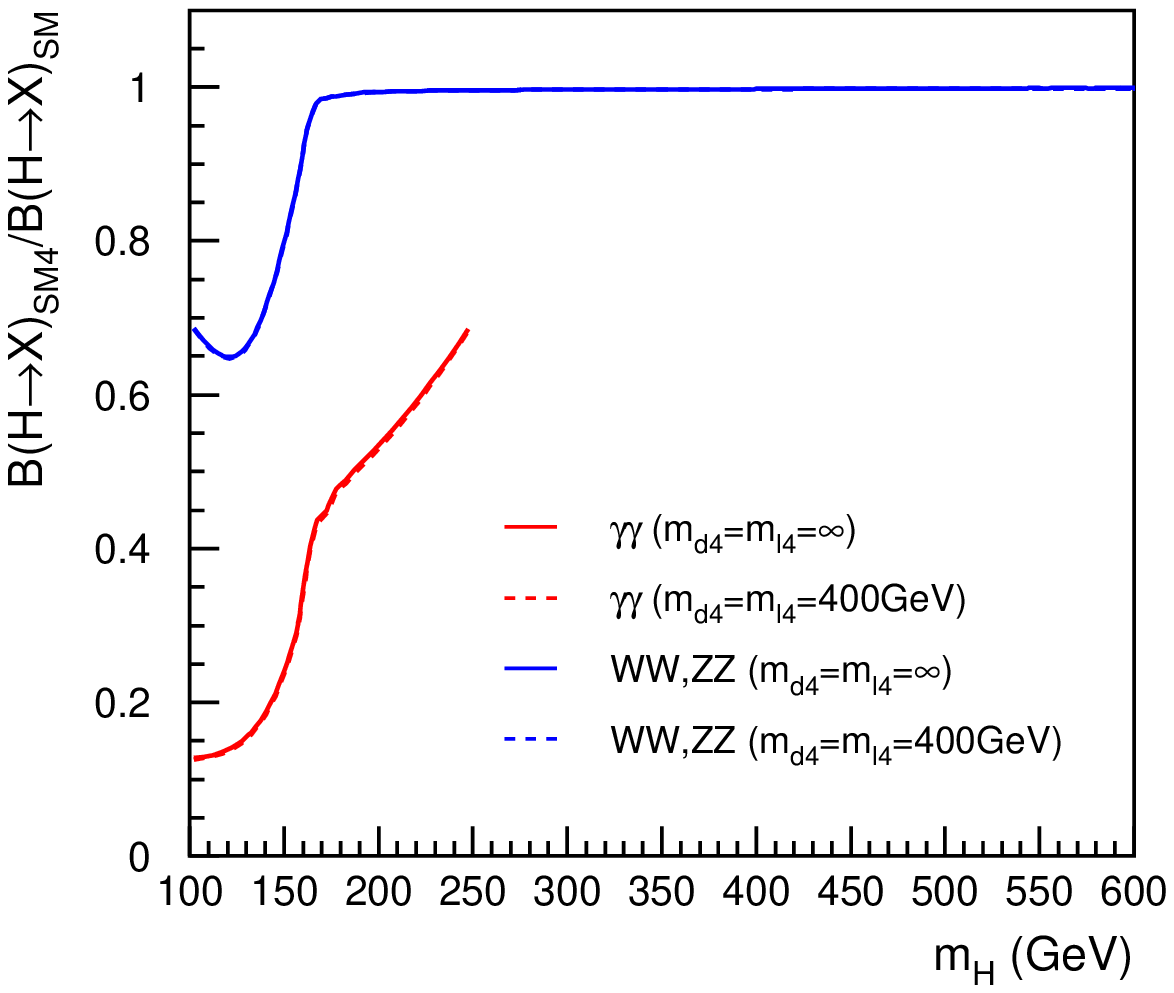}
  \includegraphics[width=0.49\columnwidth]{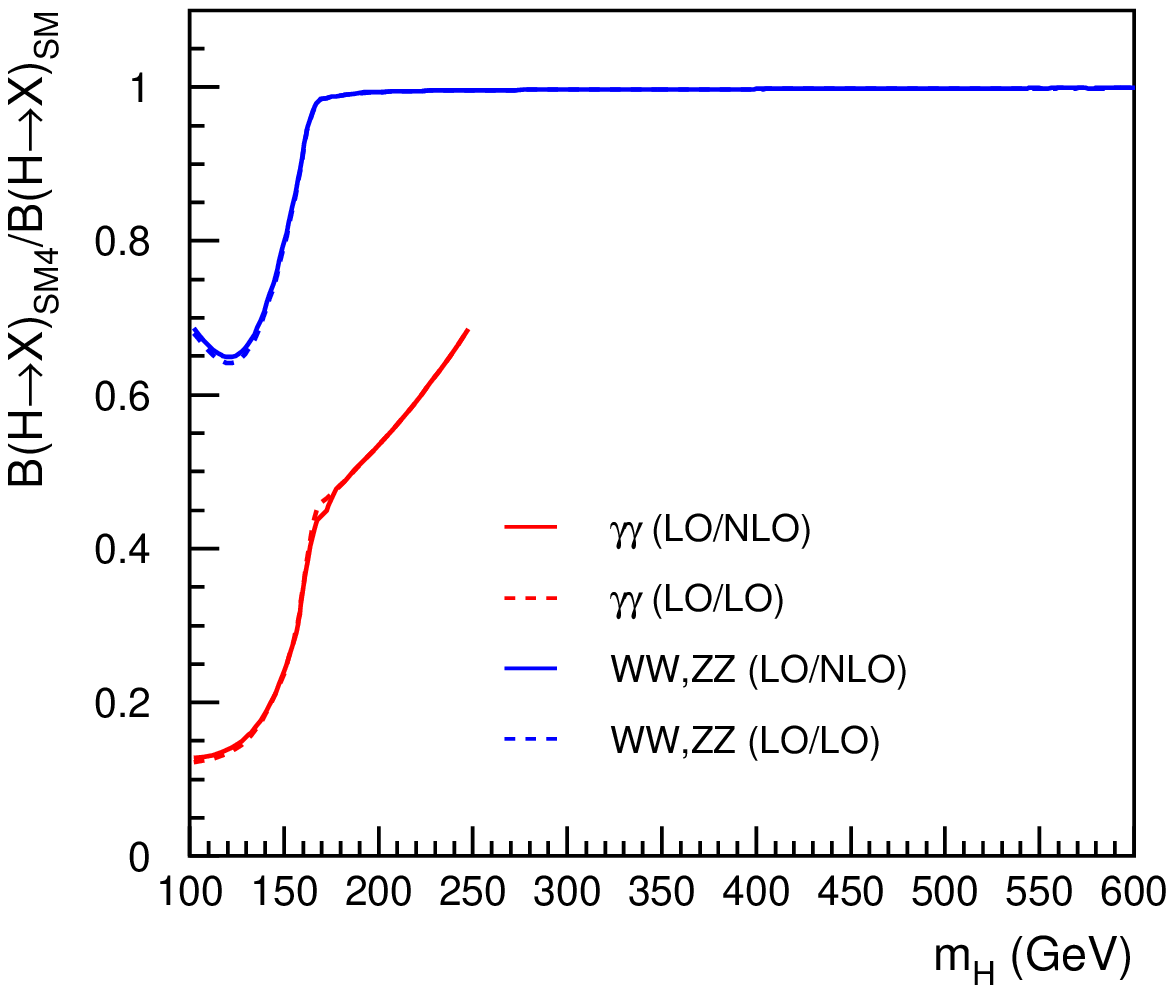}
  \end{center}
  \caption{Branching fraction ratio of Higgs decays in a fourth generation model
over that in the SM for two mass scenarios (left) and with or without 
electroweak and higher-order corrections in the SM (right).}
  \label{fig:rbr}
\end{figure}

For the relevant decay modes $\gamma\gamma$ and $WW, ZZ$, 
the ratio of the branching fractions in a fourth generation model over that 
in the SM 
\begin{equation}
R^{\rm SM4/SM}_{B(H\rightarrow X)}\equiv \frac{B(H\rightarrow X)_{\rm SM4}}{B(H\rightarrow X)_{\rm SM}}
\end{equation}
is compared in the two mass scenarios in Fig.~\ref{fig:rbr}(left),
which shows that the effect of the mass scenarios is indeed small.
The effect of the electroweak and higher-order QCD corrections on the 
branching fraction ratio is also small as it is illustrated in
Fig.~\ref{fig:rbr}(right).

\section{Results and discussions}

\begin{figure}
  \begin{center}
  \includegraphics[width=0.55\columnwidth]{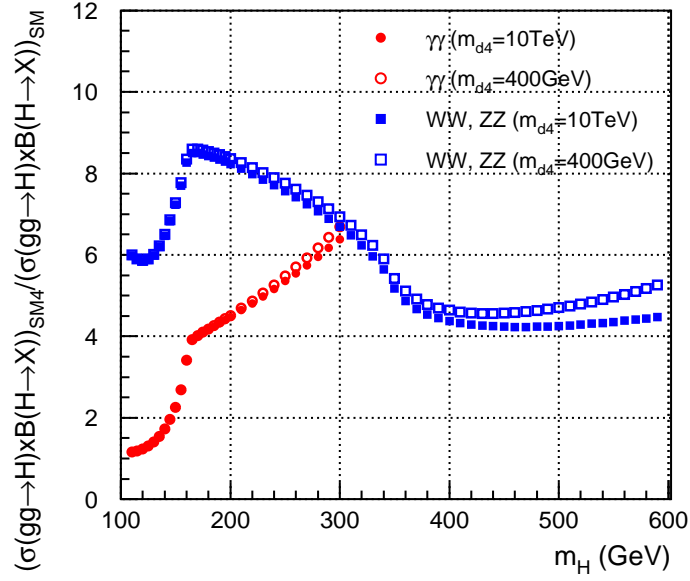}
  \end{center}
  \caption{The enhancement of the product of the cross section and the 
branching fraction in a fourth family over the SM shown as a function of 
the Higgs mass for the two mass scenarios.}
  \label{fig:xsbr}
\end{figure}

The overall enhancement of the product of the Higgs production cross section 
and the Higgs branching fraction in a fourth family over that in the SM 
\begin{equation}
R^{\rm SM4/SM}\equiv\frac{\left[\sigma(gg\rightarrow H)\times B(H\rightarrow X)\right]_{\rm SM4}}{\left[\sigma(gg\rightarrow H)\times B(H\rightarrow X)\right]_{\rm SM}}
\end{equation}
is shown in Fig.~\ref{fig:xsbr} 
as a function of the Higgs mass for decay modes $X=\gamma \gamma$, $WW$ and 
$ZZ$ and for the two mass scenarios.
The numerical values for $R^{\rm SM4/SM}_{\sigma(gg\rightarrow H)}$, $R^{\rm SM4/SM}_{B(H\rightarrow WW,\,ZZ)}$  and $R^{\rm SM4/SM}_{B(H\rightarrow \gamma\gamma)}$ for 59 
Higgs mass points ranging from 110\,GeV to 600\,GeV are given in linked web 
pages~\cite{weblink_10k,weblink_400}.

\begin{figure}
  \begin{center}
  \includegraphics[width=0.55\columnwidth]{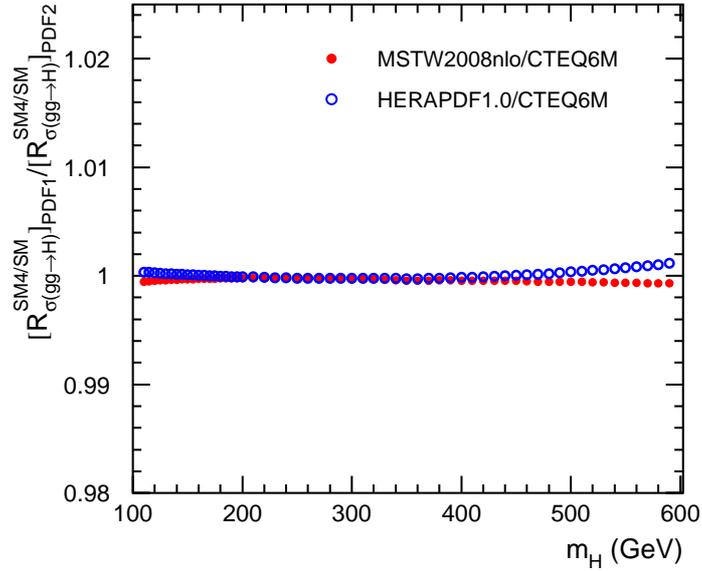}
  \end{center}
  \caption{The variation due to the choice of different PDFs on $R^{\rm SM4/SM}_{\sigma(gg\rightarrow H)}$ shown for the high mass quark scenario as a function of 
the Higgs mass $m_H$.}
  \label{fig:pdf}
\end{figure}

One advantage of reporting $R^{\rm SM4/SM}_{\sigma(gg\rightarrow H)}$ instead of 
$\sigma(gg\rightarrow H)_{\rm SM4}$ that we mentioned in Sec.~\ref{sec:sigma} 
is that the ratio is less sensitive to higher order corrections. It is also 
less sensitive to other theoretical uncertainties. One example is 
the dependence on the choice of parton distribution functions (PDFs). 
In Fig.~\ref{fig:pdf}, the variation of choosing
two different PDFs (MSTW2008NLO~\cite{mstw08} and HERAPDF1.0~\cite{herapdf}) 
has been compared with the default choice of CTEQ6M~\cite{cteq6}. 
In the considered Higgs mass range, the difference on $R^{\rm SM4/SM}_{\sigma(gg\rightarrow H)}$ is well within 0.2\%.

\section{Summary}

We have discussed the implication in
the Higgs production cross sections in the gluon-gluon fusion process and
the Higgs decay branching fractions in modes $gg$, $\gamma\gamma$ and $\gamma Z$
when including quarks and leptons of a fourth generation model.
The enhancement/modification on the gluon-gluon fusion cross section ($R^{\rm SM4/SM}_{\sigma(gg\rightarrow H)}$) and the Higgs decay branching fractions ($R^{\rm SM4/SM}_{B(H\rightarrow X)}$) of 
a fourth generation model over the SM as a function of the Higgs mass $m_H$
in the range of $100-600$\,GeV has been calculated and shown for two
mass scenarios. 
These ratios are found to have little sensitive to theoretical
variations such as the higher order corrections for both $R^{\rm SM4/SM}_{\sigma(gg\rightarrow H)}$ and $R^{\rm SM4/SM}_{B(H\rightarrow X)}$ and PDFs for $R^{\rm SM4/SM}_{\sigma(gg\rightarrow H)}$.

\section*{Acknowledgments}
The authors wish to thank M.~Spira for help and advice in using the HIGLU and
HDECAY codes. Z.~Zhang is also grateful to F.~Richard and E.~Kou for 
discussions.

%% The Appendices part is started with the command \appendix;
%% appendix sections are then done as normal sections
%% \appendix

%% \section{}
%% \label{}

%% References
%%
%% Following citation commands can be used in the body text:
%% Usage of \cite is as follows:
%%   \cite{key}         ==>>  [#]
%%   \cite[chap. 2]{key} ==>> [#, chap. 2]
%%

%% References with bibTeX database:

%\bibliographystyle{}
%\bibliography{4th}

%% Authors are advised to submit their bibtex database files. They are
%% requested to list a bibtex style file in the manuscript if they do
%% not want to use elsarticle-num.bst.

%% References without bibTeX database:

% \begin{thebibliography}{00}

%% \bibitem must have the following form:
%%   \bibitem{key}...
%%

% \bibitem{}

% \end{thebibliography}

\end{document}